\documentclass[letterpaper, 10 pt, conference]{ieeeconf}

\IEEEoverridecommandlockouts
\usepackage{graphics} % for pdf, bitmapped graphics files
\usepackage{amsmath}
\usepackage{amssymb}  % assumes amsmath package installed
\usepackage{amsfonts} % assumes amsmath package installed
\usepackage{graphicx,float,wrapfig,subfigure}
\usepackage{cite}
\usepackage{bm}
\usepackage{soul,color}
\usepackage[linesnumbered,ruled]{algorithm2e}
\usepackage{url}
\newcommand{\norm}[1]{\left\lVert#1\right\rVert}

\title{\LARGE \bf
Design and Implementation of Ecological Adaptive Cruise Control for Autonomous Driving with Communication to Traffic Lights}

\author{Sangjae Bae$^{1}$, Yeojun Kim$^{2}$, Jacopo Guanetti$^{2}$, Francesco Borrelli$^{2}$, and Scott Moura$^{1}$% <-this % stops a space
\thanks{$^{1}$Energy, Controls, and Applications Lab (eCAL), Department of Civil and Environmental Engineering, University of California, Berkeley, USA. Corresponding author: {\tt\small sangjae.bae@berkeley.edu}}%
\thanks{$^{2}$Model Predictive Control Lab (MPC lab), Department of Mechanical Engineering, University of California, Berkeley, USA.}}%
\begin{document}
\maketitle
\thispagestyle{empty}
\pagestyle{empty}

\begin{abstract}
This paper presents the design and implementation results of an ecological adaptive cruise controller (ECO-ACC) which exploits driving automation and connectivity.
The controller avoids front collisions and traffic light violations, and is designed to reduce the energy consumption of connected automated vehicles by utilizing historical and real-time signal phase and timing data of traffic lights that adapt to the current traffic conditions.
We propose an optimization-based generation of a reference velocity, and a velocity-tracking model predictive controller that avoids front collisions and violations.
We present an experimental setup encompassing the real vehicle and controller in the loop, and an environment simulator in which the traffic flow and the traffic light patterns are calibrated on real-world data.
We present and analyze simulation and experimental results, finding a significant potential for energy consumption reduction, even in the presence of traffic.
\end{abstract}

\section{Introduction}

Advanced Driving Assistance Systems (ADAS) represent the first mass deployment of driving automation technologies to the mass market.
Adaptive Cruise Control (ACC), autonomous emergency braking, and lane keeping assistance are examples of widely deployed functions in today's cars.
These technologies are already included in vehicles with so-called Level 2 automation \cite{SAE-J3016_201401}.

Driver comfort and safety are the main goals and motivations behind ADAS.
Current research in this area is focused on cooperation \cite{Plessen2016,Shen2015}, personalization \cite{Carvalho2015}, and vehicle performance \cite{AlAlam2015}.
The performance and safety gap with human drivers can be further improved via connectivity to other vehicles and to the infrastructure, which drastically increases information available to the vehicle.
%\hl{Ioannou et al. proposed an intelligent adaptive cruise control} \cite{ioannou1993autonomous}

Longitudinal control has long been studied in the control literature as a way to reduce energy consumption.
Several ACC designs have been proposed, which aim at preventing energy-wasteful behaviors, such as unnecessary braking and throttling based on the perception of the immediate surroundings \cite{Schmied2015,turri2017model}.
On a related front, finding the optimal speed trajectory from an origin to a destination, given the road topology, is a classical optimal control application that is often labeled as ``Eco-driving'' \cite{Sciarretta2015}.
In simplified settings, one can find the optimal policy analytically (e.g. the ``pulse and glide'' solution \cite{Sciarretta2015}).
In reality, real roadways impose complex constraints, notably stop signs and traffic lights, which in general require numerical methods (see e.g. \cite{Sciarretta2015, Ozatay2014, Guanetti2018}).
In particular, the presence of signalized intersections along the path introduces non-convex constraints in the optimization problem.
On the other hand, ``green waving'' through traffic lights pledges significant energy savings.
A more detailed survey of the state-of-the-art on this topic can be found at Section 5.2 of \cite{Guanetti2018}.
Our recent work \cite{sun2018robust} focused on the Eco-driving problem through signalized intersections with \emph{uncertain} effective red light duration.
%with \emph{adaptive} signalized intersections, i.e. where the red and green light durations are not fixed, but continuously adapted to the traffic conditions.
Since the corresponding optimization problem is uncertain, the crossing of intersections during green lights was formulated as a chance constraint, to be satisfied with high probability.
Then the optimization problem was parametrized using historical signal phase and timing data, and solved by Dynamic Programming (DP).
Simulations showed potential fuel savings up to 40\%, compared to a modified intelligent driver model \cite{kesting2010enhanced}.
%Kamel et al. \cite{kamal2011ecological} presented an Eco-driving control based on Model Predictive Control (MPC), especially on hilly roads. Their simulation result shows about 8\% of fuel savings of their MPC controller on a hilly road.

While there is an extensive literature on the potential energy benefits of ADAS and Eco-driving, limited experimental validation has been produced so far (mostly as enhanced cruise control systems, see e.g. \cite{Sciarretta2015}).
Since Eco-driving problems often have a long time/space horizon, only a portion of the obstacles can be considered in practice, both to reduce the problem complexity and because the actual obstacles may be unknown.
Thorough experimental validation is needed to evaluate the performance gap between the ideal setting (e.g. the vehicle driving in free flow, and no model mismatch, and the reference velocity is perfectly tracked) and the real world.
The integration of ADAS (short-sighted, but aware of traffic and sudden phase changes) and Eco-driving (long-sighted, but only aware of static or slowly changing information) is expected to play an important role in this sense, and needs to be carefully crafted.
Another challenge is that various aspects of road experiments are not reproducible, most notably the surrounding traffic.
To this end, we recently presented a vehicle-in-the-loop setup to test Connected and Automated Vehicles (CAVs) in real-world traffic conditions \cite{kim2018avec}.

The main contributions of this paper are two-fold. (i) We first propose a control framework that systematically balances energy consumption and travel time, for arterial roads with signalized intersections.
The framework includes a robust Eco-driving control, which minimizes energy consumption and incorporates effective red light duration uncertainty, and an ACC that avoids frontal collisions and obeys traffic signals.
These two controls can conflict, and therefore their integration is carefully examined. (ii) We then present a novel hardware and software setup to implement and test the online controller.
Simulations and dynamometer experiments are described which reproduce real-world signal phase and timing and traffic data from Arcadia, California.
In total, this work is comprised of algorithms with rigorous theoretical foundations in combination with a unique hardware-in-the-loop experiment for controller prototyping.

The paper is organized in the following manner.
Section \ref{sec:design_online_controller} describes the mathematical formulation of the proposed control framework.
Section \ref{sec:implementation} details the hardware and software setup to implement the control system.
Section \ref{sec:result} presents and analyzes simulation and experiments results.
Section \ref{sec:conclusion} summarizes the paper's contributions.

\section{Controller Design}
\label{sec:design_online_controller}
In this section, we described the proposed ECO-ACC controller.
The goal of this control system is to automate the longitudinal control of a vehicle driving through one or more signalized intersections.
The controller avoids front collisions, obeys traffic signals, and aims at minimizing a convex combination of energy consumption and travel time by cruising through the green phases of the various intersections.

To enhance the energy performance, the control system can access information on the Signal Phase and Timing (SPaT) of the traffic lights along the route, as well as information on the corresponding vehicle queues.
The method proposed in this paper works also when the traffic light timing is not fixed a priori, but changes depending on the traffic level (e.g. if vehicles are queued at a specific lane).
Historical data are available for both the vehicle queue lengths and the traffic lights SPaT, providing empirical probabilistic information on the variable red light durations and vehicle queue lengths. 
In real-time, the current SPaT for each traffic light can be accessed through Vehicle-to-Infrastructure (V2I) communication; moreover, the radar and camera systems provide information about the front obstacles.

In other words, the ECO-ACC controller has two horizons: (i) a \emph{slow} and \emph{long-sighted} horizon, where the scope is the overall trip and only slowly changing information (such as historical or probabilistic data) can be taken into account; and (ii) a \emph{fast} and \emph{short sighted} horizon, where the scope is the immediately upcoming portion of the trip, and dynamic information (such as real-time traffic light or obstacles) can be taken into account.

\begin{figure}
    \centering
    \includegraphics[width=1\columnwidth]{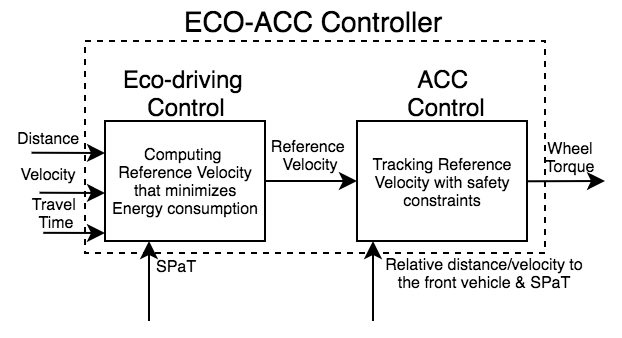}
    \caption{Diagram for the conceptual incorporation between the Eco-driving and ACC Controller}
    \label{fig:diag_incorporation_controllers}
\end{figure}

In sum, the ECO-ACC controller encompasses two components, visualized in Fig. \ref{fig:diag_incorporation_controllers}.
(i) An Eco-driving controller, which computes a reference velocity trajectory from the current location to destination, aimed at minimizing energy based on the probabilistic information on SPaT and vehicle queues.
(ii) An Adaptive Cruise Controller (ACC), which computes the wheel torque required to follow the reference velocity, yet guarantee safety -- i.e. collision avoidance and traffic signal compliance -- in spite of the presence of traffic on the road. 
This time scale separation also makes intuitive sense to tame the computational complexity of the overall control task, as we will more precisely explain in the rest of the section.

% Eco-drive controller based on chance-constrained DP
% FIGURE : layers of different controllers (cite Chao's paper)
\subsection{Eco-driving Control}
We apply the optimal Eco-driving algorithm proposed by Sun et al. in \cite{sun2018robust}.
The algorithm determines the optimal state and control trajectories to minimize fuel consumption, in the presence of uncertain red light duration.
This is mathematically formulated as a chance constrained optimal control problem, using empirical statistics on the red light duration.
For completeness we summarize the Eco-driving controller here, but readers should refer to \cite{sun2018robust} for details.

\subsubsection{Vehicle Dynamics}
Consider the longitudinal vehicle dynamics where the longitudinal acceleration $a$ is
\begin{equation}
    a = \frac{r_{gb}T_{eng}-T_{brk}}{mR_{w}}-g\left(cos(\theta)C_r -sin(\theta)\right) -\frac{\rho AC_{d}}{2m}v^2 ,
\end{equation}
where $m$ is the vehicle mass, $r_{gb}$ is the product of the gearbox and final drive ratio, $T_{eng}$ is the internal combustion engine output torque, $R_{w}$ is the wheel rolling radius, $\theta$ is road grade, $\rho$ is the air density, $A$ is the front cross-sectional area, and $C_d$ is the air drag coefficient.
The rolling resistance coefficient $C_r$ is
\begin{equation}
    C_r = C_{r1} + C_{r2}v,
\end{equation}
where $C_{r1}$ and $C_{r2}$ are constants.
We consider the velocity $v(k)$ and travel time $t(k)$ as states at position $k \Delta s$.
Consequently, the system dynamics can be written as
\begin{equation}
    \underbrace{\begin{bmatrix}
    v(k+1)\\
    t(k+1)
    \end{bmatrix}}_{x({k+1})}
    = \underbrace{\begin{bmatrix}
    v(k)\\
    t(k)
    \end{bmatrix}}_{x(k)}
    + \begin{bmatrix}
    \frac{a(k) \Delta s}{v(k)}\\
    \frac{\Delta s}{v(k) + \frac{a(k) \Delta s}{v(k)}}
    \end{bmatrix},
    \label{eq:dynamics}
\end{equation}
where $k\in\{0,\cdots,N-1\}$, $\Delta s$ is the position step size.
It is important to note that the dynamics are difference equations in space, not time.
Interested readers are referred to \cite{sun2018robust} for more details.

\subsubsection{Problem Formulation}
The objective of the online Eco-driving controller is to minimize a convex combination of the energy consumption and of the travel time, over the remaining portion of the trip, from the current position and time.
As a proxy for fuel consumption in the online implementation, we utilize squared wheel torque, where the wheel torque is written
\begin{equation}
    T_{w} = \frac{r_{gb}T_{eng}-T_{brk}}{R_{w}}.
\end{equation}
The objective function is then represented in the (discrete) spatial domain as
\begin{equation}
    J = \sum_{k \in \{0,...,N-1\}} \left(T_{w}(k)^2 + \lambda \left(\frac{1}{v(k)}\right)^2 \right)\Delta s^2, \label{eq:objective_fun}
\end{equation}
with the regularization parameter $\lambda$.
The control variable $u_k$ is the wheel torque $T_{w}(k)$, which is a weighted sum of the engine and brake torque\footnote{We do not have to separate the wheel torque into the engine and brake torques, by assuming that the subject vehicle does not simultaneously accelerate and decelerate.}.
The constraints are
\begin{align}
    T_{w}^{\text{min}} &\leq T_{w}(k) \leq T_{w}^{\text{max}}, \forall \ k\in\{0,\cdots, N\}\label{eq:ineq_whl_trq},\\
    a^{\text{min}}&\leq a(k)\leq a^{\text{max}}, \forall \ 
    k\in\{0,\cdots, N\}\label{eq:ineq_a},\\
    v^{\text{min}}&\leq v(k) \leq v^{\text{max}}(k), \forall \ k\in\{0,\cdots, N\}\label{eq:ineq_v},\\
    t(N)&\leq t_f\label{eq:ineq_t},
\end{align}
where the inequality constraints \eqref{eq:ineq_whl_trq}, \eqref{eq:ineq_a}, \eqref{eq:ineq_v}, and \eqref{eq:ineq_t} ensure the wheel torque, acceleration, velocity, and travel time, respectively, are lower and upper bounded by appropriate values, which are given and known.
Particularly, the wheel torque $T_{w}$ is lower-bounded by the maximum brake torque $T_{w}^{\min}=-\frac{T_{brk}^{\max}}{R_{w}}$, and upper-bounded by the maximum engine torque $T_{w}^{\max}=\frac{r_{gb}T_{eng}^{\max}}{R_{w}}$.
The maximum acceleration $a^{\max}$ is set to a physically feasible limit, and the maximum velocity $v^{\max}$ is set to the maximum speed limit on the road.

\subsubsection{Chance Constraints for Uncertain Effective Red Light Duration}\label{sec:chance_Constraints}
We assume perfect communication between the subject vehicle and traffic signals, i.e., the controller knows the traffic signal's current phase, timing, and cycle length.
If the cycle time and red light duration are fixed, the Eco-driving controller can ensure that the vehicle strictly obeys the traffic signals at all times, i.e. it always stops at red lights and only passes through green lights.
This can be modeled by the inequality constraints
\begin{equation}
    c_{p}^{i}(v(k),t(k),T_{w}(k)) \geq c_{r}^i(t(k)),
    \label{eq:ineq_c}
\end{equation}
where $c_{p}^{i}(v(k),t(k),T_{w}(k))$ is the cycle time at which the subject vehicle passes through the $i$-th intersection and $c_{r}^{i}(t(k))$ is the remaining red light duration at the $i$-th intersection. 
Inequality \eqref{eq:ineq_c}, however, does not model the effect of traffic queues and other unplanned delays at the intersections, which can prevent crossing the intersection even during a green light; moreover, inequality \eqref{eq:ineq_c} cannot be enforced if the red light duration is not fixed, which is the case in \emph{adaptive} traffic light controllers. % \cite{mirchandani2001real}.
To account for such cases, we introduce the random variables $\alpha^i$, i.e.
\begin{equation}
    c_{p}^{i}(v(k),t(k),T_{w}(k)) \geq c_{r}^i(t(k)) + \alpha^{i},
    \label{eq:ineq_c_a}
\end{equation}
where $\alpha^{i} \in [0,c_{g}^{i}]$ and $c_{g}^{i}$ is the green light duration at the $i$-th intersection.
The green light length $c_{g}^{i}$ is assumed to be known for all intersections.
Finally, in the optimization we enforce the chance constraints
\begin{equation}
    c_{p}^{i}(v(k),t(k),T_{w}(k)) > c_{r}^i(t(k)) + F^{-1}(1-\eta),
    \label{eq:ineq_chance}
\end{equation}
where $F^{-1}(1-\eta)$ denotes the inverse cumulative distribution function (CDF)\footnote{$F(\cdot)$ is assumed to be known and bijective.} of $\alpha^{i}$ with a desired reliability $\eta\in[0,1]$.
Interested readers are referred to \cite{sun2018robust} for details.

\subsubsection{Dynamic Programming (DP)}
We apply DP to solve the minimization problem with the objective function \eqref{eq:objective_fun},
\begin{equation}
    \min_{T_{w}\in\mathbf{R}^{N}}\;\; \sum_{k \in \{0,...,N-1\}} \left(T_{w}(k)^2 + \lambda \left(\frac{1}{v(k)}\right)^2 \right)\Delta s^2\label{eq:dp}
\end{equation}
with the system dynamics \eqref{eq:dynamics} and constraints \eqref{eq:ineq_whl_trq}-\eqref{eq:ineq_t} and \eqref{eq:ineq_chance}. Bellman's equation $V_k(x(k))$ is written
\begin{align}
    V_k(x(k)) =&\min_{T_{w}(k)} \left\{ T_{w}(k)^2 + \lambda \left(\frac{1}{v(k)}\right)^2\right.\nonumber\\
    &\quad\quad + \left.V_{k+1}(f(x(k),T_{w}(k))) \right\},\label{eq:dp_backward}
\end{align}
where $f(\cdot)$ denotes the system dynamics \eqref{eq:dynamics} with the terminal condition $V_{N}(x(N)) = T_{w}(N)^2 + \lambda\left(\frac{1}{v(N)}\right)^2$. 
Finally, solving \eqref{eq:dp_backward} backward in position step $k$ gives us the optimal wheel torque for each state.

\subsection{ACC Controller}
In this section we present our Adaptive Cruise Controller (ACC) as depicted in Fig.~\ref{fig:diag_incorporation_controllers}. 
The objective of our ACC is twofold: (i) Track the reference velocity computed by the higher level Eco-driving Controller. (ii) Enforces vehicle safety, such as collision avoidance with a leading vehicle, and non-violation of traffic light laws regardless of the front vehicle's future behavior and traffic light dynamics.
To enforce safety at all times, we use robust model predictive control for our ACC. This approach has been shown effective in longitudinal vehicle dynamics control \cite{lefevre2016learning}.

\subsubsection{Simplified Longitudinal Vehicle Dynamics}\label{sec:acc_veh_dynamics}
In ACC controller, we use the vehicle dynamics based on time step $t$ instead of the dynamics based on position step $k$ \eqref{eq:dynamics}.
The states of our model, $x$, include distance to the upcoming traffic light $d_{TL}$, the distance to the front vehicle $d_f$, and the vehicle velocity $v$. 
The input to the model, $u$, is the wheel torque $T_{w}$.
At time step $t$, the system dynamics discretized with the sampling time $t_s$ is expressed as
\begin{equation}\label{eq:vehicle_model_time}
    \underbrace{\begin{bmatrix}
    d_{TL}(t+1)\\
    d_{f}(t+1)\\
    v(t+1)
    \end{bmatrix}}_{x({t+1})}
    = \underbrace{\begin{bmatrix}
   d_{TL}(t) - t_s v(t)\\
   d_f(t) + t_s (v_{f}(t) - v(t))\\
   v(t) + \frac{t_s}{m} ( \frac{T_{w}(t)}{R_{w}} - T_{R}(t))
    \end{bmatrix}}_{f(x(t), u(t), v_{f}(t))},
\end{equation}
where $v_f$ is the front vehicle velocity and $T_{R}$ is the resistance torque approximated as
\begin{align} \label{eq:resistance_force}
T_{R}(t) = m g C_r  + \frac{1}{2}\rho A C_d v(t)^2,
\end{align}
where we assume that the road is flat at all times.
Moreover, we denote by $a^{\text{max}}_{\text{dec}}$ and $t^{max}_{\text{stop}}$ the maximum deceleration which the vehicle is capable of at any time and the maximum time required for the vehicle to come to a full stop from any initial velocity, respectively. They can be expressed as
\begin{align} \label{eq:min_dec}
a^{\text{max}}_{\text{dec}} & =\frac{T_{w}^{\text{min}}}{mR_{w}}  - g C_r, \\
t^{max}_{\text{stop}} & = -\frac{v^{\text{max}}}{a^{\text{max}}_{\text{dec}}}.
\end{align}

\subsubsection{Safety Constraints}\label{sec:acc_safety_constraints}
The safety constraints are related to avoiding collision with the front vehicle, stopping at the red light, and obeying speed limits of the road. 
% The constraint to ensure stopping at the red light is only enforced if the following condition is met: the upcoming traffic is red and the minimum stopping distance is larger than $d_{\text{TL}}$. At time $t$, this condition can be expressed as 
% \begin{align}
%  p^{\text{up}}(t) &= \text{red}, \label{eq:tl_condition1} \\
%  -\frac{v(t)^2}{2a^{\text{min}}_{\text{dec}}} &\geq d_{\text{TL}}(t)\,,\label{eq:tl_condition2} 
% \end{align}
% where $p^{\text{up}}(t)$ denotes the phase of the upcoming traffic light at time step $t$. 
They can be written together as
% \begin{align}
% & d^\text{min} \leq d_f (t) \label{eq:collision_kafety}\,,\\
% &\text{if  } p^{\text{up}}(t) = \text{red},  \nonumber \\
% & \quad 0 \leq d_{TL} (t) ,
% \label{eq:TL_safety}\\
% &\text{if  } p^{\text{up}}(t) = \text{yellow},  \nonumber \\
% & \quad 0 \leq d_{TL} (t) + \phi(t), \quad \phi(t) \geq 0 \, , \label{eq:TL_yel_safety}\\
% &v^\text{min} \leq v (t) \leq v^\text{max}\,.
% \end{align}
\begin{align}
& d^\text{min} \leq d_f (t) \label{eq:collision_kafety}\,,\\
& \quad 0 \leq d_{TL} (t) ,\quad\quad\quad
 \text{ if  } p^{\text{up}}(t) = \text{red}, \label{eq:TL_safety}\\
& \quad 0 \leq d_{TL} (t) + \phi(t), \text{ if  } p^{\text{up}}(t) = \text{yellow},\label{eq:TL_yel_safety}\\
&v^\text{min} \leq v (t) \leq v^\text{max},
\end{align}
where $p^{\text{up}}(t)$ denotes the phase of the upcoming traffic light at time step $t$; $\phi(t)$ denotes the slack variable, $\phi(t)\geq0$. Here, we assume that the yellow light phase is long enough for the vehicle to come to a full stop with the maximum deceleration \eqref{eq:min_dec}; i.e., $t^{\text{yellow}} \geq t^{max}_{\text{stop}}$.
This assumption with the soft constraint in \eqref{eq:TL_yel_safety} ensures that the vehicle is either capable of the full stop before the traffic light or passes the light when the traffic light turns to red from yellow.
% ; therefore, the constraint \eqref{eq:TL_safety} is satisfied when the traffic light changes from yellow to red.

\subsubsection{Problem Formulation}\label{sec:acc_prob_formulation}
In our ACC we compute the optimal input trajectories $u^{\ast}(\cdot|t)$ by solving at time $t$ the following finite time-horizon optimal control problem:
\begin{subequations}
\begin{align}
\underset{u(\cdot|t)}{\text{min}} \quad &J = 
\sum_{\ell=t}^{t+N_p} \norm{v (\ell | t) - v_\text{ref}}_{W_v} \label{eq:trackingpenalty}\\
 &\;\;\,+\sum_{\ell=t}^{t+N_p-1} \norm{u (\ell | t)}_{W_u} \label{eq:inputpenalty}\\
&\;\;\,+\sum_{\ell=t+1}^{t+N_p-1} \norm{u (\ell | t)- u (\ell-1 | t)}_{W_{\Delta u}} \label{eq:jerkpenalty}\\
&\;\;\,+\sum_{\ell=t+1}^{t+N_p-1} \norm{\phi (\ell | t)}_{W_{\phi}} \label{eq:slackpenalty}\\
\text{subject to} \quad 
&x(\ell+1|t) = f(x(\ell|t),\, u(\ell|t),\, v_f(\ell\mid t))\,,\nonumber\\
& d^\text{min} \leq d_f (\ell\mid t) \,, \nonumber\\
& \begin{cases}
0 \leq d_{TL} (\ell\mid t)&\text{if  } p^{\text{up}}(t) = \text{red,}\\
0 \leq d_{TL} (\ell\mid t) + \phi(\ell\mid t)&\text{if  } p^{\text{up}}(t) = \text{yellow,}
\end{cases}\nonumber\\
&v^\text{min} \leq v (\ell\mid t) \leq v^\text{max}\, \nonumber\\
&T_{w}^{\text{min}} \leq T_{w}(\ell\mid t) \leq T_{w}^{\text{max}}\,, \nonumber\\
&x(t|t) = x(t)\,, \nonumber\\
&\begin{cases}
    x(t+N_p|t) \in \mathbb{C}_{\text{TL}} \,& \text{if  } p^{\text{up}}(t) = \text{red},\\
    x(t+N_p|t) \in \mathbb{C}_{f} & \text{otherwise,}
\end{cases}\label{eq:terminalset}
% &x(\ell+1|t) = f(x(\ell|t),\, u(\ell|t),\, v_f(\ell\mid t))\,,\nonumber\\
% & d^\text{min} \leq d_f (\ell\mid t) \,, \nonumber\\
% &\text{if  } p^{\text{up}}(t) = \text{red},  \nonumber \\
% & \quad 0 \leq d_{TL} (\ell\mid t) ,
% \nonumber\\
% &\text{if  } p^{\text{up}}(t) = \text{yellow},  \nonumber \\
% & \quad 0 \leq d_{TL} (t) + \phi(t), \quad \phi(t) \geq 0 \, ,\nonumber \\
% &v^\text{min} \leq v (\ell\mid t) \leq v^\text{max}\, \nonumber\\
% &T_{w}^{\text{min}} \leq T_{w}(\ell\mid t) \leq T_{w}^{\text{max}}\,, \nonumber\\
% &\forall \ell=t,...,t+N_p-1, \nonumber \\
% &x(t|t) = x(t)\,, \nonumber\\
% &\text{if  } p^{\text{up}}(t) = \text{red},\nonumber \\
% & \quad x(t+N_p|t) \in \mathbb{C}_{\text{TL}} \,,  \label{eq:terminalset} \\
% &\text{else,} \nonumber\\
% & \quad x(t+N_p|t) \in \mathbb{C}_{f} \label{eq:TLterminalset}
\end{align}
\label{eq:mpcproblem}
\end{subequations}
for all $\ell\in\{t,...,t+N_p-1\}$ where $x(\ell\mid t)$ and $u(\ell\mid t)$ are the predicted states and input at time $\ell$ based on the measurements and predictions at time $t$, respectively; $N_p$ denotes the prediction horizon of our problem; $v_\text{ref}$ is the desired optimal velocity reference obtained from the wheel torque of the Eco-driving controller and the dynamics \eqref{eq:vehicle_model_time}. 
The cost function $J$ includes a penalty for deviating from $v_\text{ref}$ \eqref{eq:trackingpenalty}, a penalty on input torques \eqref{eq:inputpenalty}, a penalty for jerk \eqref{eq:jerkpenalty}, and, finally, a penalty for violating the soft yellow light constraint \eqref{eq:slackpenalty}.
The weights for each penalty are $W_v$, $W_u$, $W_{\Delta u}$, and $W_{\phi}$, respectively.
The polytopic constraint \eqref{eq:terminalset} enforces the terminal state to lie inside the \textit{robust control invariant} sets defined in \cite{lefevre2016learning} for recursive feasibility of our finite horizon optimal control problem. 
To construct these sets, we assume the linear uncertain version of our dynamics model \eqref{eq:vehicle_model_time} by considering the nonlinear air drag term, $\frac{1}{2}\rho A C_d v(t)^2$, as a linear addictive and bounded uncertainty.
% Moreover, for $\mathbb{C}_{\text{TL}}$ in \eqref{eq:TLterminalset}, we assume there is a static (non-moving) object at the upcoming traffic light.
The first input $u^{*}_{0|t}$ is applied to the system during the time interval $[t,t+1)$ and at the next time step $t+1$, a finite horizon optimal control problem \eqref{eq:mpcproblem} with new state measurements, is solved over a shifted horizon, yielding a \textit{moving} or \textit{receding} horizon control strategy.

\subsection{Integration of Eco-driving control and ACC}
The Eco-driving controller can conflict with the ACC, in the sense that the ACC can cause the vehicle to deviate from the optimal velocity profile computed by the Eco-driving controller.
The ACC takes into account the actual traffic around the vehicle and robust avoidance of front collisions.
The Eco-driving controller only accounts for the long-term traffic signal timing and essentially assumes free-flow conditions\footnote{Even though the historical data of vehicle queue length is considered, this information is essentially modeled as a longer red light.}.
However, if we removed the Eco-driving controller and just used ACC (with constant velocity tracking), this would lead to a less energy efficient behavior, as shown below.
In order to mitigate this issue, the ACC Controller also penalizes jerk (i.e. the derivative of acceleration), as a trade-off with velocity tracking.
This yields smoother reference velocity tracking in the presence of surrounding traffic.
We tune the weights in \eqref{eq:inputpenalty} and \eqref{eq:jerkpenalty} based on experimental results to achieve a balance of smooth reference velocity tracking and collision avoidance.

\section{Hardware in the loop setup}
\label{sec:implementation}
\begin{figure}
    \centering
    \includegraphics[width=1\columnwidth]{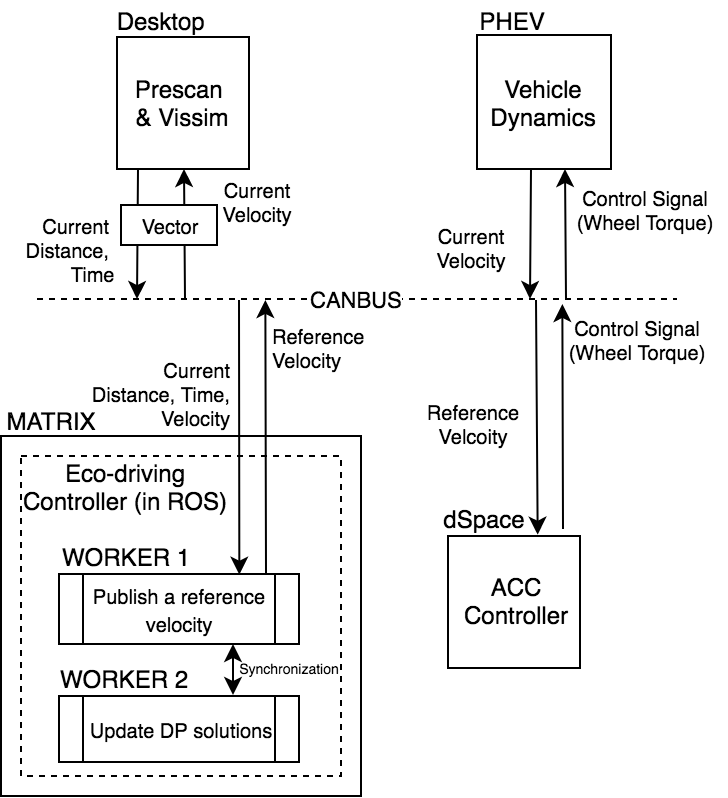}
    \caption{Diagram for the implementation setup of the ECO-ACC controller}
    \label{fig:diag_implement_environment}
\end{figure}

Our setup consists of four main components communicating over CAN bus, as depicted in Fig.~\ref{fig:diag_implement_environment}: (i) the subject vehicle, a Plug-in Hybrid Electric Vehicle (PHEV), placed on a dynamometer; (ii) a desktop computer, running the simulator of the environment surrounding the vehicle; (iii) a dSPACE MicroAutoBox II, running the ACC software; (iv) an Adlink Matrix embedded PC, running the Eco-driving control software.
% \begin{enumerate}
%     \item the subject vehicle, which for this study is a plug-in hybrid electric vehicle (PHEV), and is placed on a dynamometer;
%     \item a desktop computer, running the simulator of the environment surrounding the vehicle, as describd below;
%     \item a dSPACE MicroAutoBox II, running the ACC software;
%     \item an Adlink Matrix embedded PC, running the Eco-driving control software.
% \end{enumerate}
The basic specifications of the above components are provided in Table \ref{table:1}.
Readers are referred to \cite{kim2018avec} for more details.

\begin{table}[]
\begin{tabular}{|c|c|}
\hline
Item & Specifications \\ \hline
Desktop & \begin{tabular}[c]{@{}c@{}}Intel(R) Core(TM) i7-7700KK CPU @\\ 4.20Hz with NVIDIA GeForce GTX 1080\end{tabular} \\ \hline
\begin{tabular}[c]{@{}c@{}}Matrix embedded \\ PC-Adlink\end{tabular} & \begin{tabular}[c]{@{}c@{}}MXC-6101D/M4G with Intel Core\\ i7-620LE 2.0 GHz processor\end{tabular} \\ \hline
\begin{tabular}[c]{@{}c@{}}dSpace \\ MicroAutoBox\end{tabular} & IBM PowerPC 750FX processor, 800 MHz \\ \hline \begin{tabular}[c]{@{}c@{}}Plug-in Hybrid\\ Electric Vehicle\end{tabular} & \begin{tabular}[c]{@{}c@{}}8.89 kWh of battery capacity\end{tabular}\\ \hline
\end{tabular}
\caption{Specifications of hardware and software in the implementation setup}
\label{table:1}
\end{table}

\subsection{Environment model}

The environment model, including the traffic lights and the road and intersections geometry, is constructed using PreScan \cite{Prescan}.
The density and velocity of the other vehicles on the road in the PreScan environment is determined by PTV Vissim \cite{Vissim}, a state-of-the-art traffic microsimulator.
PreScan and Vissim interact in such a way that the motion of vehicles in PreScan respects the trajectories generated by Vissim, and the trajectories in Vissim are adapted to the motion of the subject vehicle as simulated in PreScan.

\begin{figure}
    \centering
    \includegraphics[width=1\columnwidth]{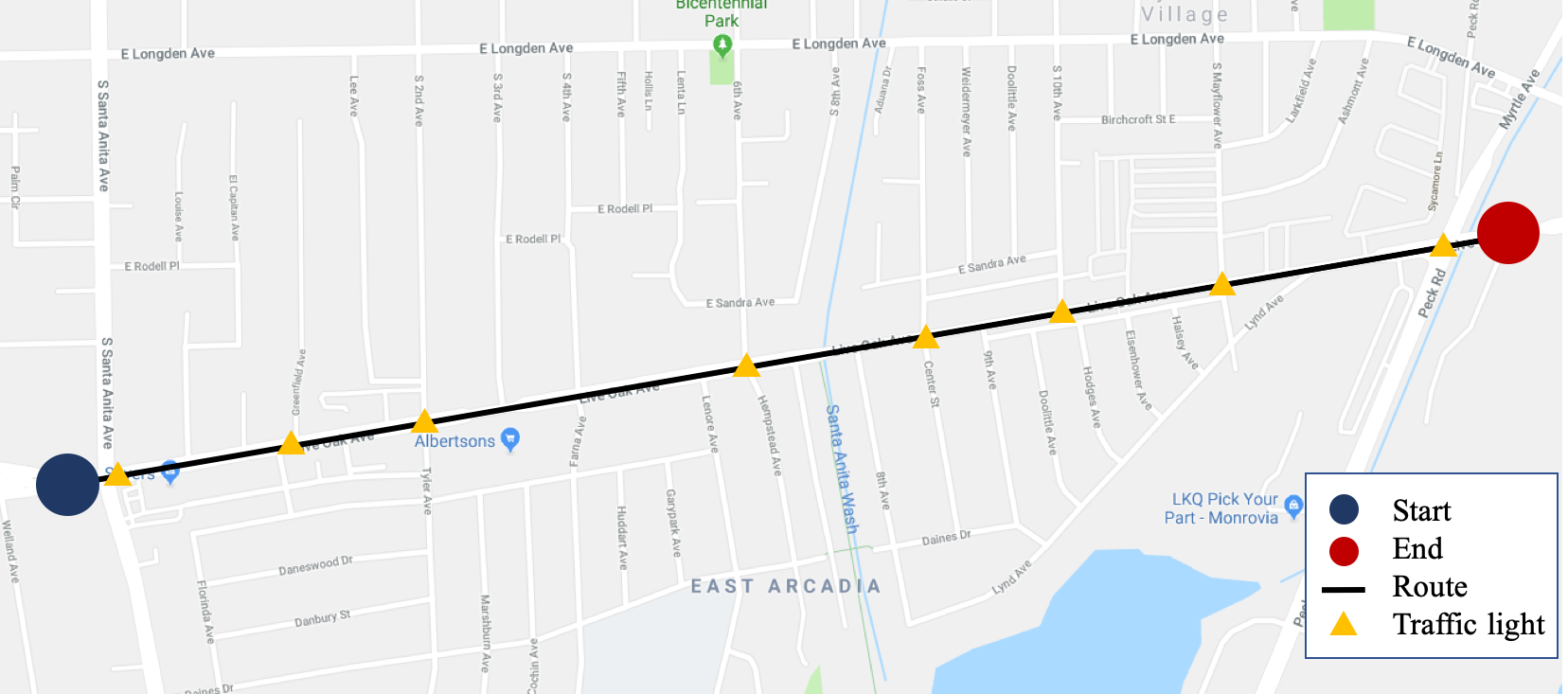}
    \caption{Test route on the Live Oak corridor in Arcadia, CA, USA. The 8 signalized intersections, marked by yellow triangles, are located respectively at $\{42, 351, 610, 1190, 1509, 1764, 2050, 2456\}$ meter from the route origin.}%located respectively at 42 m, 351 m, 610 m, 1190 m, 1509 m, 1764 m, 2050 m, 2456 m from the route origin.}
    \label{fig:arcadia}
\end{figure}

\begin{figure}
    \centering
    \includegraphics[width=1\columnwidth]{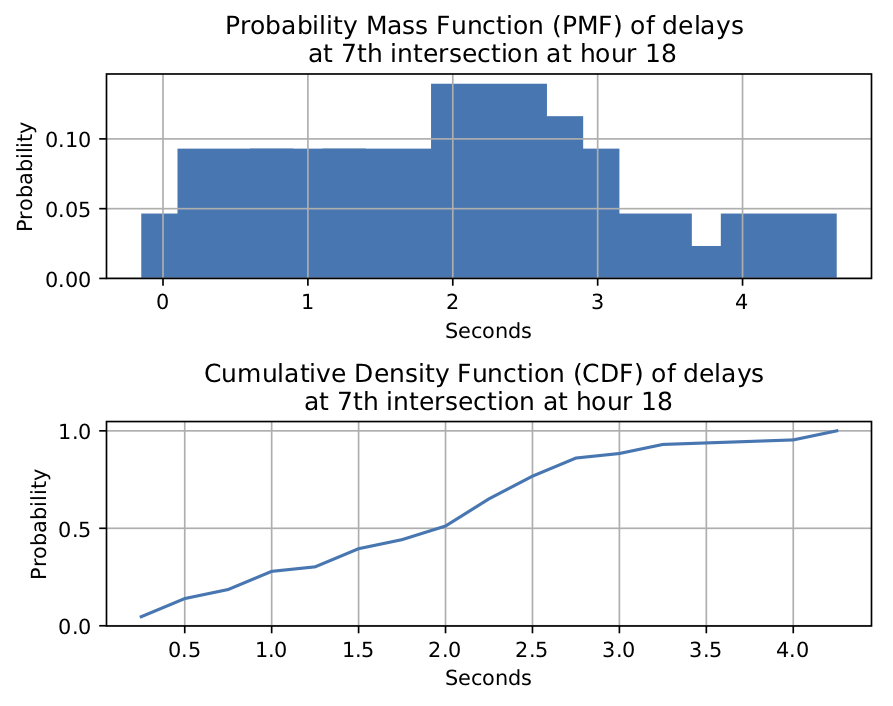}
    \caption{Empirical PMF (top) and CDF (bottom) of delays in passing through the $7^{th}$ intersection (from the origin) after the traffic signal turns to green, resulted by traffic queues.}
    \label{fig:pdf}
\end{figure}

In this paper, we consider a scenario of automated driving along the Live Oak Avenue in Arcadia, CA, USA.
The driving route, depicted in Fig.~\ref{fig:arcadia}, is 2.6 km long and has 8 signalized intersections.
The Vissim model takes the route inflows and outflows as inputs; these parameters were measured on the actual road, by means of vehicle detectors.
At the intersections, the vehicles generated by the Vissim model can either turn or continue along Live Oak Avenue, according to a probabilistic turn policy; such policy is also parametrized by the recorded aggregate traffic volume and turn counts data \cite{kim2018avec}.
Because no vehicle-to-vehicle communication is assumed, the behavior of the surrounding vehicles, e.g. decelerations or lane changes, are unpredictable for the subject vehicle. Such uncertainty is dealt with by the ACC system. The Signal Phase and Timing (SPaT) data of the traffic lights are input to both PreScan and Vissim, as they affect both the subject vehicle and the traffic.
For each intersection, we first compute an empirical probability mass function (PMF) based on the corresponding historical SPaT data, collected over the month of July 2018.
For the experiments presented in this paper, the traffic light patterns are sampled from such functions.
The corresponding Cumulative Density Function (CDF) is used to parametrize the chance constraints~\eqref{eq:ineq_chance}.
Fig.~\ref{fig:pdf} shows the PMF and CDF for the $7^{th}$ intersection from the origin, as an example; at 6 pm at this intersection, vehicles wait 1.96 seconds on average with a standard deviation of 1.033 seconds.

\subsection{Controller implementation}

The Eco-driving controller implements two concurring tasks.
One task updates continuously the DP solution solving problem~\eqref{eq:dp}, %solving the backward iteration~\eqref{eq:dp_backward} 
from the current state until the end of the trip; 
The other task, every 200 ms, outputs a reference velocity for the ACC; this is obtained querying, with the current state, the optimal control policy computed by the most recent DP run.
The Eco-driving controller is implemented in ROS and the computations are performed calling Matlab; we utilize the Matlab Parallel Computing Toolbox \cite{matlabparallel} to assign the update of the DP solution to one worker and the update of the reference velocity to another worker. The MPC-based ACC is implemented solving the nonlinear optimization problem~\eqref{eq:mpcproblem} every 200 ms with NPSOL \cite{gill1986user}.

\section{Implementation Result}
\label{sec:result}
In this section, we present and discuss the results of an experiment performed on the setup described above.
In particular, we compare the performance of the ECO-ACC controller to a baseline: the ACC controller without the Eco-driving controller (denoted as ACC-Only controller). 
In the implementation of the baseline controller (ACC-Only), we set the reference velocity to a constant 15 m/s, which is the speed limit along the route.

\subsection{Energy Performance of the ECO-ACC Controller}
\begin{figure} 
    \centering
    \includegraphics[width=1\columnwidth]{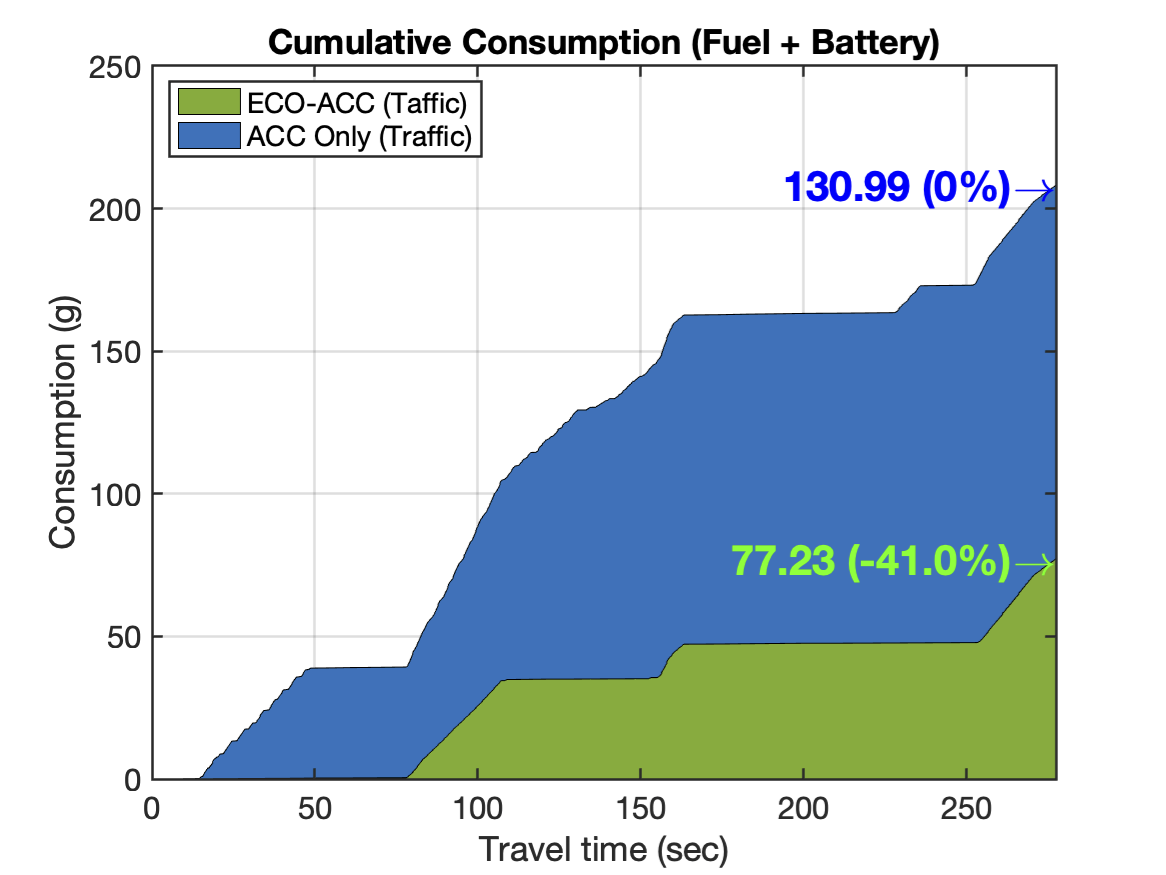}
    \caption{Area chart for the cumulative equivalent fuel consumption. We calculate that 33.7 kilowatt-hours of electricity is equivalent to one gallon of gas \cite{friedman2014analysis}. The values with percentage in parentheses indicate the relative reduction of equivalent fuel consumption compared to that of the ACC controller without the Eco-driving controller.}
    \label{fig:fuel}
\end{figure}

Fig.~\ref{fig:fuel} shows the cumulative equivalent fuel consumption over the travel time as an area chart.%, where the height of each area represents the value of cumulative equivalent fuel consumption.
The equivalent fuel consumption evaluates both gas and battery consumption, as explained in the caption of Fig.~\ref{fig:fuel}.
The ECO-ACC controller significantly reduces ($41.0\%$) the equivalent fuel consumption compared to the ACC, for the given route and traffic; this improvement can be attributed to the use of the SPaT information by the Eco-driving controller.
%In other words, the controller has global information regarding the traffic signals. In practice, there would be a limited distance over which the vehicle can access traffic signal SPaT information, which we further discuss in \ref{sec:limitations}.

\begin{figure}
    \centering
    \includegraphics[width=\columnwidth]{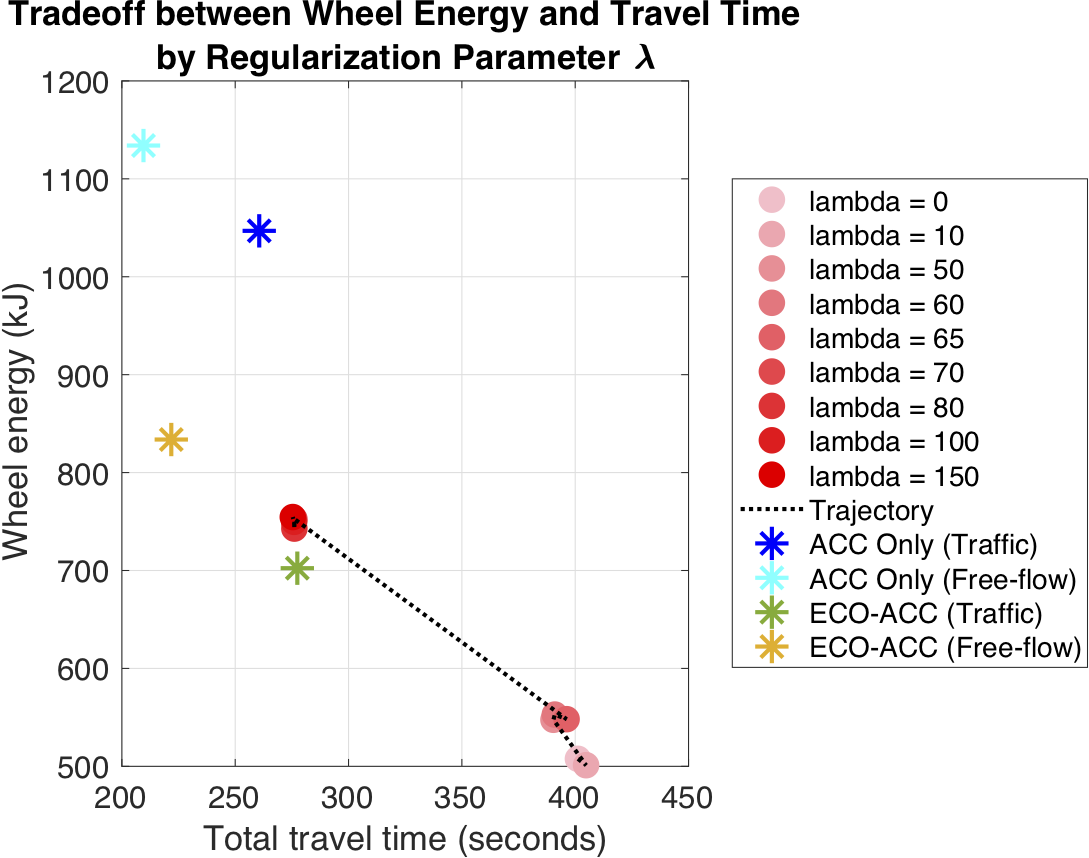}
    \caption{Pareto chart of regularization parameter $\lambda$ of the ECO-ACC controller simulation with traffic. The maximum travel time is set to 400 seconds. The points of circle shape indicate simulation results with a mathematical vehicle model and those of asterisk shape indicate implementation results with a real PHEV.}
    \label{fig:pareto}
\end{figure}

Because of the presence of traffic on the road, the travel time along the route cannot be exactly fixed for two different experiments; as a consequence, the energy performance must be evaluated along with the travel time.
Fig.~\ref{fig:pareto} illustrates the trade-off between the total wheel energy and the travel time for various experiments and simulations.
The stars in green and dark blue show the performance of the ECO-ACC and ACC-Only controllers in the experiments just discussed.
The ECO-ACC controller uses 0.1951 (kWh) of wheel energy, which is 32.91\% less than the ACC-Only controller; on the other hand, the ACC-Only controller takes 6\% less travel time (260.6 seconds) than the ECO-ACC controller (277.4 seconds).

In general, the trade-off between travel time and energy consumption depends on the traffic and can be empirically adjusted, in the Eco-driving controller, by tuning the parameter $\lambda$ in \eqref{eq:objective_fun}.
To investigate this trade-off, we analyze experiments and simulations, also summarized in Fig.~\ref{fig:pareto}. The stars in yellow and light blue represent the performance of the ECO-ACC and ACC-Only controllers in another set of experiments, conducted in the same conditions but without traffic; we observe similar gaps between the two controllers, for both energy consumption and travel time. The red dots represent the performance of the ECO-ACC controller obtained in simulation, under the same traffic and traffic light conditions used for the experiments; different points correspond to different values of $\lambda$. The dotted line simply provides a visual grouping, and does not represent a continuous relationship between wheel energy and total travel time. As expected, the wheel energy tends to increase as $\lambda$ (i.e., the penalty on the total travel time) increases.
%This is intuitive as the controller decides to apply more wheel torque in order to arrive at the destination sooner. 
We observe two clusters of red points, i.e. there is a ``jump" in performance for a small variation of $\lambda$ (from $\lambda=65$ to $\lambda=70$); this is because there are fixed, discrete time windows where the vehicle can pass through the intersections.
In Fig.~\ref{fig:speed_traveltime}, the time windows are illustrated as time gaps between the red dashes.
We also notice that the wheel energy obtained from the ECO-ACC controller with the the PHEV is slightly lower than that obtained with a mathematical vehicle model.
We conjecture the simulations over-predict wheel energy consumption due to model mismatch.
Nevertheless, the experimental performance of the ECO-ACC controller seems fairly close to the optimal Pareto curve found in simulation, whereas the ACC-Only controller seems significantly suboptimal.

\subsection{Wheel Torque and SOC Profile}
\begin{figure}
    \centering
    \includegraphics[width=1\columnwidth]{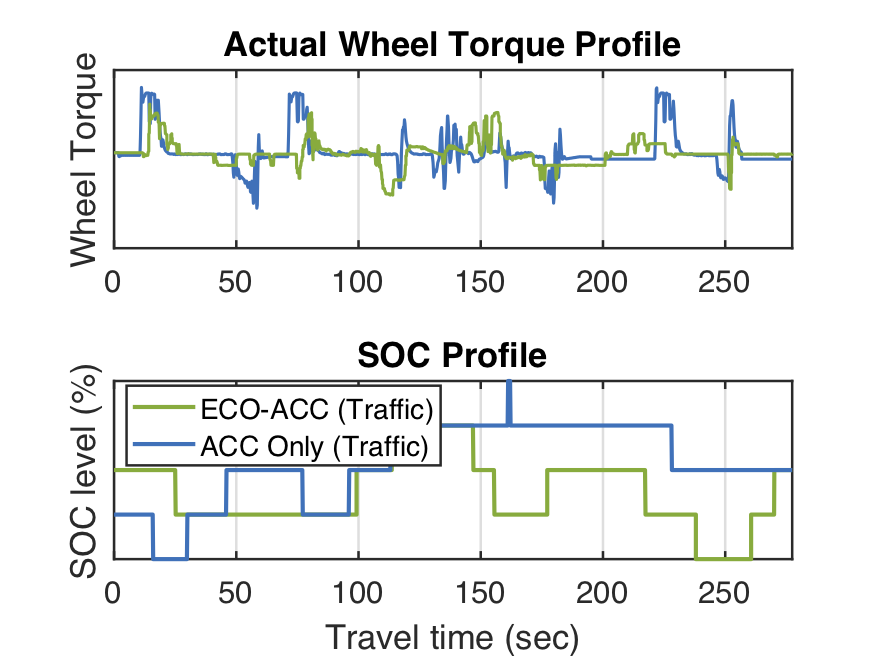}
    \caption{Torque (top) and SOC (bottom) profiles. The initial conditions of SOC differ by 0.5\%. The vertical scales of both plots have been omitted for confidentiality reasons.}
    \label{fig:torque}
\end{figure}

The top plot in Fig.~\ref{fig:torque} shows the wheel torque profile applied on vehicle.
The trajectory is more aggressive in the ACC-Only case, which tries to keep a constant 15 m/s velocity, ignoring the future traffic lights.
This is consistent with the higher wheel energy consumption observed in Fig.~\ref{fig:pareto}.
The bottom plot of Fig.~\ref{fig:torque} shows that the SOC level remains almost constant throughout the trip for both controllers (the steps correspond to 0.5\% variations).
In terms of equivalent fuel, the battery energy accounts for 0.618 (g) in the ECO-ACC case and for 0.556 (g) for the ACC-Only case; consequently, battery energy consumption plays a relatively small role in this experiment.% in comparing the ECO-ACC and ACC-Only controllers.

\subsection{Velocity Profile and Travel Time}

\begin{figure}
    \centering
    \includegraphics[width=1\columnwidth,trim={1cm 1cm 0.5cm 1cm},clip]{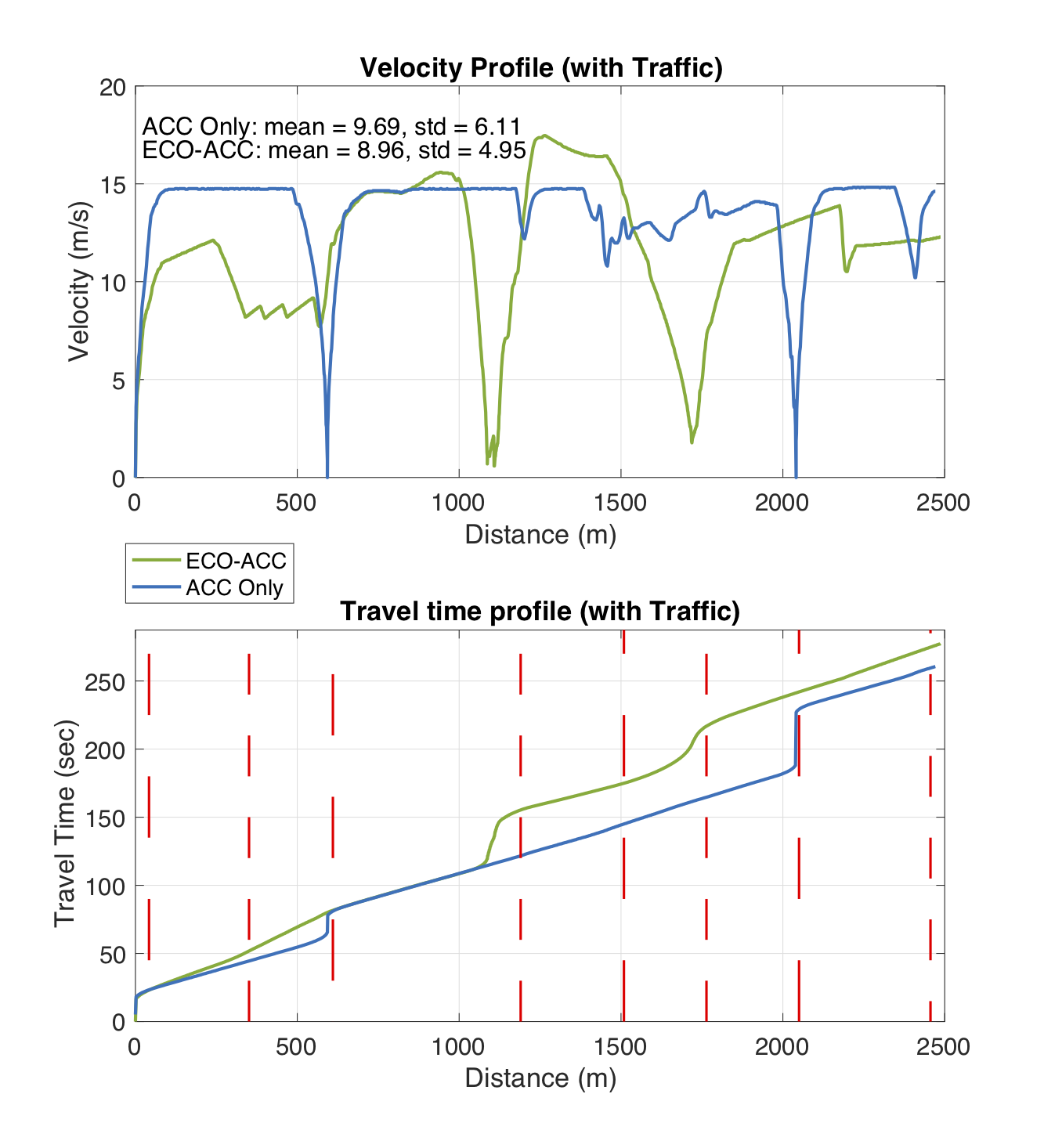}
    \caption{Velocity (top) and travel time (bottom) profiles over travel distance. The red dashed lines represent the red light periods. The chattering in the velocity profiles occurs as the ACC controller balances speed with distance from the front vehicle.}
    \label{fig:speed_traveltime}
\end{figure}

Fig.~\ref{fig:speed_traveltime} shows the speed and travel time profiles over travel distance.
The velocity profile of the ECO-ACC controller never gets to zero once the trip starts, although it gets close to zero (0.762 m/s).
In other words, the vehicle does not stop at any red light from the start to the end of the trip.
The bottom plot in the same figure verifies that the ECO-ACC controller passes through all the intersections without having to stop at the red lights. 
We also found that the mean and standard deviation of speed are reduced by 0.73 m/s and 1.16 m/s, respectively with the ECO-ACC controller. %Nevertheless, the ACC-Only controller takes 6\% less travel time (260.6 seconds) than the ECO-ACC controller (277.4 seconds). This result can be adjusted by tuning the regularization parameter $\lambda$ in \eqref{eq:objective_fun}. 

\subsection{Impact of Traffic on Velocity Profile}
\begin{figure} 
    \centering
    \includegraphics[width=1\columnwidth]{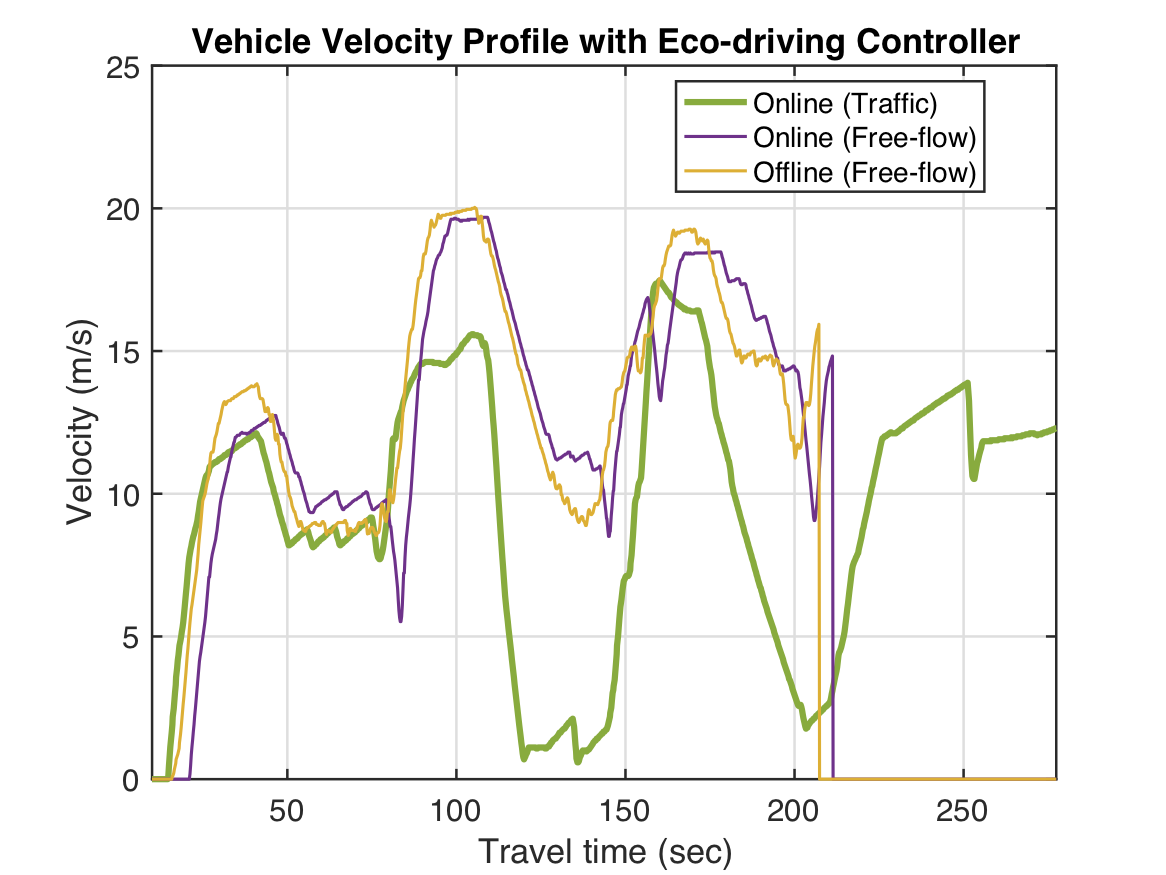}
    \caption{Comparison in vehicle velocity profiles. The velocity profile of the online controller without traffic, i.e., free-flow, is similar to that of the offline controller without traffic.}
    \label{fig:veh_Vel}
\end{figure}

We further investigate how the velocity profile varies if there is no traffic, i.e., free-flow. Fig.~\ref{fig:veh_Vel} shows the velocity profiles of the ACC controller with three types ECO-ACC controllers: (i) online planning with traffic (green); (ii) online planning without traffic (purple); and (iii) offline planning without traffic (orange). The online controllers solve the Eco-driving optimal control problem as the vehicle drives and receives updated SPaT information. The offline case computes the Eco-driving control solution once prior to the trip start, and does not recalculate it again. In free-flow conditions (no traffic), the online and offline velocity trajectories are very similar. However, with traffic, the velocity profile deviates significantly due to delays caused by other vehicles cutting in the lane or driving slowly. Consequently, the trip takes an additional 50 seconds more. As a side note, we also implemented the case of an offline ECO-ACC controller with traffic. The trip, however, did not succeed and the vehicle ended up stranded in the middle of the route. That is because the offline Eco-driving plan is not robust to unexpected delays from neighboring vehicles. This, in fact, provides excellent motivation for integrating Eco-driving with adaptive cruise control \emph{real-time}. Their combination is necessary to achieve the safety, energy economy, and quality of service we desire.

\subsection{Limitations \& Future works}\label{sec:limitations}
There exist some limitations to the proposed control designs and implementation. One limitation is that the Eco-driving controller provides a spatially-varying control policy for the entire trip, which could be computationally intractable for long trips. Moreover, in practice a vehicle will only be able to receive information from upcoming traffic signals that are a limited distance away. A receding horizon controller can be used to address these issue. Additionally, the proposed Eco-driving controller exclusively plans a velocity profile on a single lane. An extension of this work can incorporate multilane driving \cite{ajanovic2017energy}) and left/right turns. 

\section{Conclusion}\label{sec:conclusion}
This paper provides a formalized mathematical control design and experimental implementation of an online Ecological Adaptive Cruise (ECO-ACC) control architecture. Its overall objective is to minimize energy consumption while avoiding collisions and complying with traffic signals. Our proposed control architecture consists of the two levels of real-time controllers; Eco-driving controller and ACC controller. In the higher level, the Eco-driving controller computes the energy-optimal velocity reference incorporating stochastic red light duration delays via chance-constrained optimal control. In the lower level, the ACC controller ensures safety against collision with front vehicles and obeying traffic light rules. Our control design is experimentally validated through a recently developed hardware-in-the-loop setup, which consists of a Plug-in Hybrid Electric Vehicle. Case studies are performed along a 2.6 km route with eight signalized intersections. The environment and traffic is calibrated to mimic Live Oak Avenue in Arcadia, CA, USA. The experiments reveal a fuel savings 41\% for the ECO-ACC controller, relative to a ACC-Only controller. Future work includes extending the controller with a receding distance horizon formulation, to address limited communication distances and computation.

% Appendices should appear before the acknowledgment.

\section*{Acknowledgement}
The information, data, or work presented herein was funded in part by the Advanced Research Projects Agency-Energy (ARPA-E), U.S. Department of Energy, under Award Number DE-AR0000791. The views and opinions of authors expressed herein do not necessarily state or reflect those of the United States Government or any agency thereof.

The authors would also like to thank Hyundai America Technical Center, Inc. for providing us with the vehicle and a testing facility, and Sensys Networks for granting access to the traffic data used in this paper.

% \vfill

% \bibliographystyle{ieeetr}
\bibliography{ref}

\begin{thebibliography}{10}

\bibitem{SAE-J3016_201401}
{Taxonomy and Definitions for Terms Related to On-Road Motor Vehicle Automated
  Driving Systems}, 2014.

\bibitem{Plessen2016}
Mogens~Graf Plessen, Daniele Bernardini, Hasan Esen, and Alberto Bemporad.
\newblock {Multi-automated vehicle coordination using decoupled prioritized
  path planning for multi-lane one- and bi-directional traffic flow control}.
\newblock In {\em 55th IEEE Conference on Decision and Control}, pages
  1582--1588, 2016.

\bibitem{Shen2015}
Xiaotong Shen, Zhuang~Jie Chong, Scott Pendleton, Wei Liu, Baoxing Qin, Guo
  Ming~James Fu, and Marcelo~H. Ang.
\newblock {Multi-vehicle motion coordination using V2V communication}.
\newblock In {\em IEEE Intelligent Vehicles Symposium}, pages 1334--1341, 2015.

\bibitem{Carvalho2015}
Ashwin Carvalho, St{\'{e}}phanie Lef{\'{e}}vre, Georg Schildbach, Jason Kong,
  and Francesco Borrelli.
\newblock {Automated driving: The role of forecasts and uncertainty - A control
  perspective}.
\newblock {\em European Journal of Control}, 24:14--32, 2015.

\bibitem{AlAlam2015}
Assad Alam, Bart Besselink, Valerio Turri, Jonas M{\aa}rtensson, and
  Karl~Henrik Johansson.
\newblock {Heavy-duty vehicle platooning for sustainable freight
  transportation: A cooperative method to enhance safety and efficiency}.
\newblock {\em IEEE Control Systems Magazine}, 35(6):34--56, 2015.

\bibitem{Schmied2015}
Roman Schmied, Harald Waschl, and Luigi {Del Re}.
\newblock {Extension and experimental validation of fuel efficient predictive
  adaptive cruise control}.
\newblock In {\em American Control Conference}, volume 2015-July, pages
  4753--4758, 2015.

\bibitem{turri2017model}
Valerio Turri, Yeojun Kim, Jacopo Guanetti, Karl~H Johansson, and Francesco
  Borrelli.
\newblock A model predictive controller for non-cooperative eco-platooning.
\newblock In {\em American Control Conference (ACC), 2017}, pages 2309--2314.
  IEEE, 2017.

\bibitem{Sciarretta2015}
Antonio Sciarretta, Giovanni {De Nunzio}, and Luis~Leon Ojeda.
\newblock {Optimal Ecodriving Control: Energy-Efficient Driving of Road
  Vehicles as an Optimal Control Problem}.
\newblock {\em IEEE Control Systems Magazine}, 35(5):71--90, 2015.

\bibitem{Ozatay2014}
Engin Ozatay, Simona Onori, James Wollaeger, Umit Ozguner, Giorgio Rizzoni,
  Dimitar Filev, John Michelini, and Stefano {Di Cairano}.
\newblock {Cloud-based velocity profile optimization for everyday driving: A
  dynamic-programming-based solution}.
\newblock {\em IEEE Transactions on Intelligent Transportation Systems},
  15(6):2491--2505, 2014.

\bibitem{Guanetti2018}
Jacopo Guanetti, Yeojun Kim, and Francesco Borrelli.
\newblock {Control of Connected and Automated Vehicles: State of the Art and
  Future Challenges}.
\newblock {\em Annual Reviews in Control}, 45:18--40, 2018.

\bibitem{sun2018robust}
Chao Sun, Jacopo Guanetti, Francesco Borrelli, and Scott Moura.
\newblock Robust eco-driving control of autonomous vehicles connected to
  traffic lights.
\newblock {\em arXiv preprint arXiv:1802.05815}, 2018.

\bibitem{kesting2010enhanced}
Arne Kesting, Martin Treiber, and Dirk Helbing.
\newblock Enhanced intelligent driver model to access the impact of driving
  strategies on traffic capacity.
\newblock {\em Philosophical Transactions of the Royal Society of London A:
  Mathematical, Physical and Engineering Sciences}, 368(1928):4585--4605, 2010.

\bibitem{kim2018avec}
Yeojun Kim, Samuel Tay, Jacopo Guanetti, and Francesco Borrelli.
\newblock Hardware-in-the-loop for connected automated vehicles testing in real
  traffic.
\newblock In {\em 14th International Symposium on Advanced Vehicle Control (to
  appear)}, 2018.

\bibitem{lefevre2016learning}
St{\'e}phanie Lefevre, Ashwin Carvalho, and Francesco Borrelli.
\newblock A learning-based framework for velocity control in autonomous
  driving.
\newblock {\em IEEE Transactions on Automation Science and Engineering},
  13(1):32--42, 2016.

\bibitem{Prescan}
Prescan, ver:8.1.0 [computer software].
\newblock \url{https://tass.plm.automation.siemens.com/prescan}.

\bibitem{Vissim}
Vissim, ver:08 [computer software].
\newblock \url{http://vision-traffic.ptvgroup.com/en-us/products/ptv-vissim}.

\bibitem{matlabparallel}
Matlab parallel computing toolbox [computer software].
\newblock \url{https://www.mathworks.com/help/distcomp/index.html}.

\bibitem{gill1986user}
Philip~E Gill, Walter Murray, Michael~A Saunders, and Margaret~H Wright.
\newblock User's guide for npsol (version 4.0): A fortran package for nonlinear
  programming.
\newblock Technical report, Stanford Univ CA Systems Optimization Lab, 1986.

\bibitem{friedman2014analysis}
Peter~D Friedman and Phil Grossweiler.
\newblock An analysis of us federal mileage ratings for plug-in hybrid electric
  vehicles.
\newblock {\em Energy Policy}, 74:697--702, 2014.

\bibitem{ajanovic2017energy}
Zlatan Ajanovic, Michael Stolz, and Martin Horn.
\newblock Energy efficient driving in dynamic environment: considering other
  traffic participants and overtaking possibility.
\newblock In {\em Comprehensive Energy Management--Eco Routing \& Velocity
  Profiles}. Springer International Publishing AG, 2017.

\end{thebibliography}
%\bibliography{survey}

%\addtolength{\textheight}{-6cm}  % This command serves to balance the column lengths
                                  % on the last page of the document manually. It shortens
                                  % the textheight of the last page by a suitable amount.
                                  % This command does not take effect until the next page
                                  % so it should come on the page before the last. Make
                                  % sure that you do not shorten the textheight too much.

\vfill

\end{document}